\documentclass{aa}  
\usepackage{graphicx}
\usepackage{txfonts}
\usepackage{natbib,twoopt}
\bibpunct{(}{)}{;}{a}{}{,}             %% natbib format for A&A and ApJ
\usepackage{hyperref}
\hypersetup{
    colorlinks=true,
    linkcolor=blue,
    citecolor = blue,
    filecolor=magenta,
    urlcolor=blue}
\usepackage{booktabs}

\begin{document}

\authorrunning{Fanelli et al.}
\titlerunning{High resolution H-band spectroscopy of Liller~1 giant stars}

\title{Multi-iron subpopulations in Liller~1 from high resolution H-band spectroscopy
\thanks{Based on observations collected at the W. M. Keck Observatory, which is operated as a scientific partnership among the California Institute of Technology, the University of California, and the National Aeronautics and Space Administration. The Observatory was made possible by the generous financial support of the W. M. Keck Foundation.}}

\author{C. Fanelli\inst{1},
    L. Origlia\inst{1}
    R. M. Rich\inst{2},
    F. R. Ferraro\inst{1}\fnmsep\inst{3},
    D. A. Alvarez Garay\inst{1}\fnmsep\inst{3},
    L. Chiappino\inst{1}\fnmsep\inst{3},
    B. Lanzoni\inst{1}\fnmsep\inst{3},
    C. Pallanca\inst{1}\fnmsep\inst{3},
    C. Crociati\inst{1}\fnmsep\inst{3} and 
    E. Dalessandro\inst{1},
    }

   \institute{INAF, Osservatorio di Astrofisica e Scienza dello Spazio di Bologna, Via Gobetti 93/3, I-40129 Bologna, Italy
    \email{cristiano.fanelli@inaf.it}
         \and
         Department of Physics and Astronomy, UCLA, 430 Portola Plaza, Box 951547, Los Angeles, CA 90095-1547, USA
         \and
         Dipartimento di Fisica e Astronomia, Università degli Studi di Bologna, Via Gobetti 93/2, I-40129 Bologna, Italy
         }

\abstract
{We present a high resolution chemical study of a representative sample of 21 luminous giant stars of Liller~1, a complex stellar system in the Galactic bulge, based on H band spectra acquired with the Near InfraRed Spectrograph at KeckII.
15 stars turn out to have a subsolar iron abundance and enhanced [$\alpha$/Fe] and [Al/Fe], likely old that formed early and quickly from gas mainly enriched by type~II supernovae, and 6 stars with supersolar iron and roughly solar-scaled [$\alpha$/Fe] and [Al/Fe], likely younger, thus formed at later epochs from gas also enriched by type~Ia supernovae. Moreover, both subpopulations show enhanced [N/Fe], as in the bulge field, about solar-scaled [V/Fe], and depletion of [C/Fe] and $^{12}$C/$^{13}$C with respect to the solar values, indicating the occurrence of significant mixing in the stellar interiors of these evolved stars.
The current study has also made evident that the sub-solar subpopulation shows some structuring, and the presence of a third subcomponent with iron content and [$\alpha$/Fe] enhancement somewhat intermediate between the metal-poor and metal-rich main subpopulations, has been statistically assessed, providing the chemical signature of an extended star formation with multiple bursts and of some self-enrichment.}
\maketitle
\section{Introduction} 
\label{intro}
Star clusters in the Galactic bulge usually are $\sim 12$ Gyr old stellar systems with masses in the 10$^4$ $\lesssim$ M/M$_{\odot}$ $\lesssim$ 10$^6$ range, and with overall evolutionary and chemical properties typical of genuine globular clusters, that is single-age and single-iron populations with some spread in the light element abundances \citep[see e.g.][and references therein]{carretta09,carretta10,caloi11,milone17,villanova19,cadelano23}.
Notable exceptions are Terzan~5 and Liller~1, two stellar systems of the bulge with present-day masses that exceed 10$^6$ M$_{\odot}$ \citep{Lanzoni10, ferraro_21} that host stellar subpopulations with a significant spread in age and metallicity.
In particular, they both host a main stellar component with an old age (11-12 Gyr), $\alpha$-enhanced chemical mixture, and sub-solar iron abundance ($-1.0 <$ [Fe/H] $<0$ in Terzan 5, and $-0.5 <$ [Fe/H] $<0$ in Liller 1), cohabiting with 
one or more stellar populations that are super-solar, about solar-scaled alpha, and significantly younger, with an age of 4-5 Gyr in Terzan~5, and 6-8 Gyr and 1-3 Gyr in Liller~1 \citep{Ferraro_09, ferraro_16,Origlia_11,Origlia_13,origlia_19,Massari_14,ferraro_21,dalessandro_22,crociati_23,deimer24}.
In the past years, a number of scenarios for the formation and evolution of these peculiar systems have been proposed, although none of them can reasonably explain all the chemical, kinematic and evolutionary properties 
derived so far from the state-of-the-art photometric and spectroscopic observations mentioned above.
In particular, an extra-Galactic origin of Terzan~5 and Liller~1 \citep[see e.g,][]{brown18,alfaro19,taylor22} has been almost discarded based on kinematic \citep[see][]{massari_15,mas19,baum19} and age-metallicity \cite[see e.g.][and references therein]{pfeffer21} considerations, that favour an in-situ formation and evolution within the Galactic bulge. 
Metallicity also makes it poorly plausible an halo, metal-poor Galactic environment for the formation of Liller~1 and Terzan~5 \citep{moreno22}.
Also scenarios that foresee the merger of two globulars \citep{khoperskov18,mastrobuono19,pfeffer21} cannot account for the multi-age subpopulations of Terzan~5 and Liller~1, while the accretion of a giant molecular cloud by a genuine globular cluster \citep{mckenzie18,bastian_22} 
seems at variance with the observed iron distributions, which show multiple peaks and, at least in Liller~1, also provides evidence of an underlying continuous star formation \citep{dalessandro_22}. 
Based on the fact that an in situ bulge formation best accounts for the observed kinematic and metallicity of Liller~1 and Terzan~5, in order to simultaneously explain their multi-age and multi-iron distributions as well as their complex chemistry, our group made the hypothesis that these systems could be fossil fragments of the pristine clumps of stars and gas that could have contributed to forming the early bulge \citep[e.g,][and references therein]{immeli_04, elemgreen_08}.
These fragments could have survived complete disruption and evolved as independent stellar systems within the bulge, possibly self-enriching \citep{romano23} and experiencing multiple events of star formation \citep{dalessandro_22}.

This paper presents a detailed chemical study of the Liller~1 stellar populations based on high spectral resolution (R$\approx$25,000) spectroscopic observations  with the Near InfraRed Spectrograph (NIRSpec) at KeckII in the H-band, thus complementing the study of \citet[][hereafter AG24]{deimer24} based on medium resolution spectroscopy (R$\approx$8,000) in the near-IR (NIR) J,H,K bands with X-shooter at the Very Large Telescope (VLT).
NIR observations are critical to study the Liller~1 stellar populations, due to its huge absolute and differential reddening \citep{Pallanca_21} that makes observations in the optical from difficult to impossible.
Observations and spectral analysis, including target selection and membership assessment via proper motions (PMs), are described in Sect.~\ref{obs}, results for the radial velocities (RVs) and chemical abundances are presented in Sect.~\ref{results}, while in Sect.~\ref{disc} we present some discussion of the chemical constraints to the formation and evolution of Liller~1 and finally in Sect.~\ref{conclu} we draw our conclusions.
    \begin{table}
    \scriptsize
    \renewcommand{\arraystretch}{1.25}
    \setlength{\tabcolsep}{13pt}
    \caption{Observed stars in Liller 1.}
    \label{tab1}
    \begin{tabular}{|c|c|c|c|c|}
    \hline\hline
    ID &  RA   &  Dec  &   J    &   K    \\
    \hline
       & [Deg] & [Deg] & [mag]  & [mag]  \\
    \hline
    1   &  263.3165200 &  -33.3746570 & 10.890 & 7.539   \\
    3   &  263.3247780 &  -33.3730850 & 11.018 & 7.924   \\
    5   &  263.3527951 &  -33.3878338 & 11.221 & 8.499  \\
    6   &  263.3753680 &  -33.3720930 & 11.530 & 8.554   \\
    12  &  263.3534273 &  -33.3881756 & 11.623 & 8.914  \\
    14  &  263.3447000 &  -33.3890570 & 11.764 & 8.973  \\
    20*  &  263.3413160 &  -33.3917580 & 12.448 & 9.223  \\
    23  &  263.3495745 &  -33.3896670 & 11.943 & 9.136  \\
    24*  &  263.3580030 &  -33.3844380 & 12.018 & 9.257  \\
    27*  &  263.3463770 &  -33.3867450 & 12.111 & 9.273  \\
    31*  &  263.3386050 &  -33.3953320 & 12.147 & 9.315  \\
    35*  &  263.3561440 &  -33.3858760 & 12.117 & 9.276  \\
    37*  &  263.3408550 &  -33.3894200 & 12.251 & 9.377  \\
    39* &  263.3619960 &  -33.3893320 & 12.160 & 9.241  \\
    41  &  263.3510700 &  -33.3840680 & 12.282 & 9.476  \\
    %45  &  263.3540437 &  -33.3894655 & 12.020 & 9.480  \\
    57  &  263.3827190 &  -33.3711430 & 12.701 & 9.714   \\
    63  &  263.3397190 &  -33.3879780 & 12.611 & 9.822  \\
    67  &  263.3871830 &  -33.3880000 & 12.804 & 9.804   \\
    70  &  263.3700460 &  -33.3805470 & 12.853 & 9.853   \\
    74*  &  263.3490706 &  -33.3910950 & 12.874 & 10.040 \\
   % 89  &  263.3766080 &  -33.3861240 & 12.827 & 10.108  \\ %no PM
    91  &  263.3659880 &  -33.3803560 & 12.958 & 10.110 \\
    120* &  263.3599690 &  -33.3839110 & 13.128 & 10.424 \\
    \hline\hline
    \end{tabular}
    
    \vspace{0.15cm}
    $^*$Stars in common with AG24.\\
    \end{table}
\begin{figure}
    \centering
    \includegraphics[width=\columnwidth]{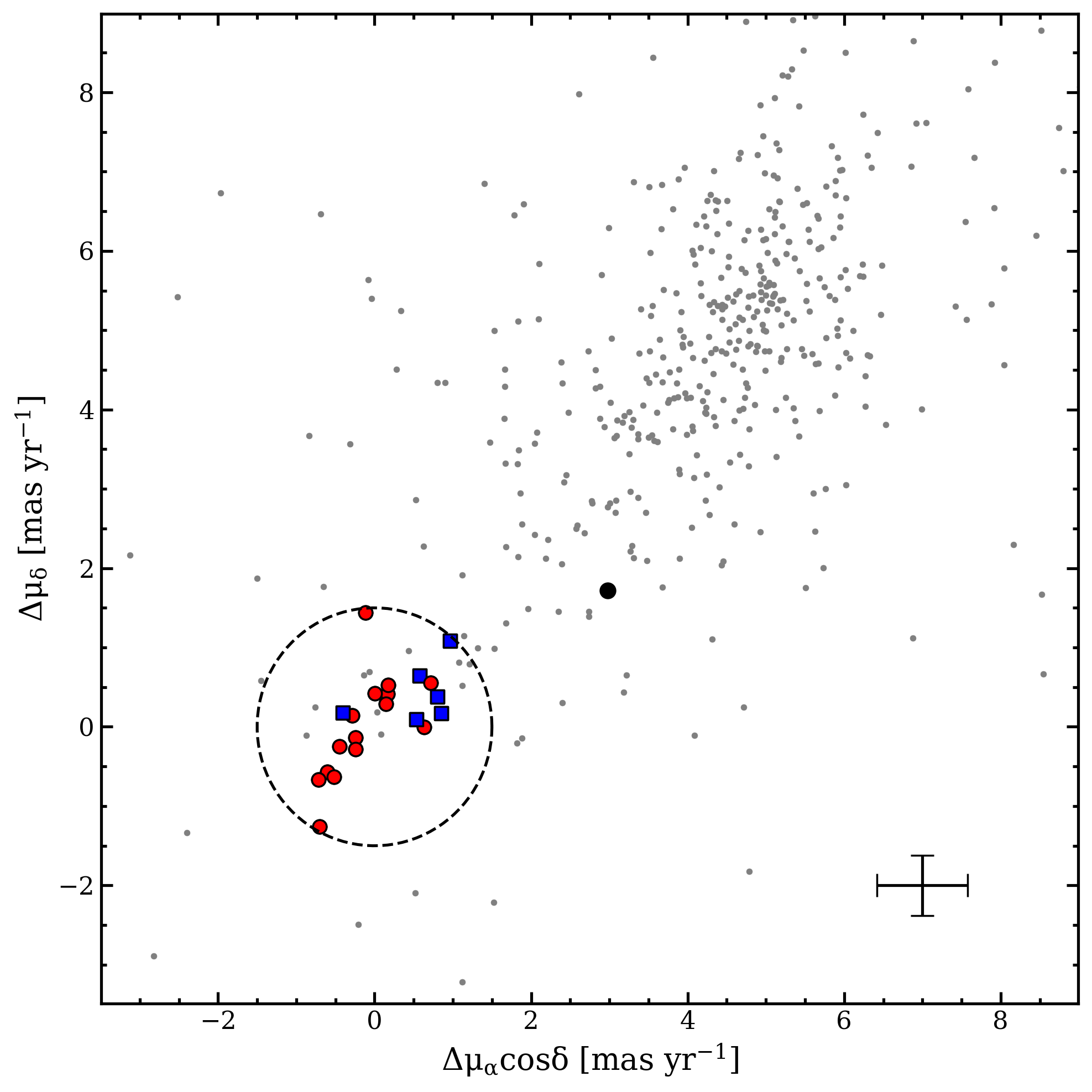}
    \caption{\textit{Gaia} DR3 PMs referred to the systemic values of Liller 1 \citep[from][]{vasiliev_21}, for the stars with magnitude comparable to that of the spectroscopic targets ($G<19$) and located within 2 arcmin from the centre of the system (gray dots) where the 22 spectroscopic targets (coloured filled circles and squares) are also located. The red filled circles and blue filled squares are likely member stars of Liller~1 belonging, respectively, to the metal-poor and  the metal-rich subpopulations (see Sect.~\ref{abu}). The typical error bars for the spectroscopic targets are reported in the bottom-right corner. The large dashed circle has a radius equal to 3 times the PM dispersion of Liller~1 (see text). The spectroscopic target located beyond the dashed circle (star \#57, marked as a black filled circle) likely is a Galactic field interloper and has been excluded from the chemical analysis.} 
\label{fig1}
\end{figure}

\begin{figure*}
%\centering
    \includegraphics[width=\textwidth]{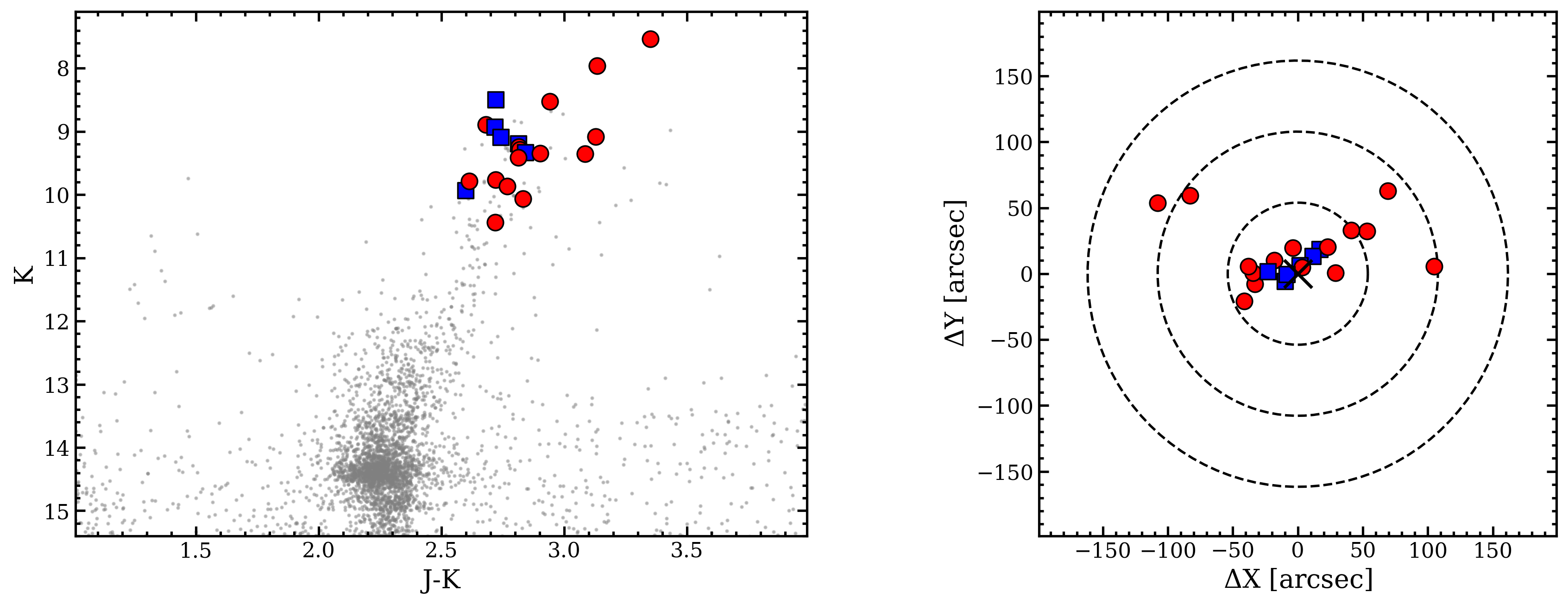}
    \caption{Liller~1 K, J-K  CMD (left panel, gray dots) with highlighted the metal-poor (red filled circles) and metal-rich (blue filled squares) likely member stars, for which we measured chemical abundances (see Sect.~\ref{abu}), and distribution of these stars on the plane of the sky with respect to its center, marked with the black cross and located at RA=$263\rlap{.}^\circ3523333$, Dec=$-33\rlap{.}^\circ3895556$ (right panel). The dashed black circles have radii equal to 10, 20, and 30 times the core radius $r_c = 5\rlap{.}^{"}39$ \citep{saracino_15}.}
    \label{fig2}
\end{figure*}

\section{Observations and spectral analysis}
\label{obs}
We acquired high resolution spectra of 22 giant stars toward Liller~1 with the NIRSpec echelle spectrograph \citep{mclean98} at Keck II. Observations were undertaken on 2012, April 28 and on 2013, May 13 and June 17. 
We used the NIRSpec-5 setting to enable observations in the H-band and a 0.43" slit width that provided an overall spectral resolution of R = 25,000.
Data reduction was performed by using the REDSPEC IDL-based package developed at the UCLA IR Laboratory.
Each spectrum was sky subtracted by using nod pairs, corrected for flat-field and calibrated in wavelength using arc lamps. 
An O-star spectrum observed during the same night was used to identify and remove telluric features. 
The signal-to-noise ratio per resolution element of the final spectra is always $\ge$40. 

Table~\ref{tab1} lists the observed target stars, their coordinates and J and K magnitudes from our compilation of NIR photometry \citep{valenti10,ferraro_21} and from the VISTA Variables in the Via Lactea (VVV, \citealt{minniti_10}). Nine stars are in common with AG24. 
Before proceeding with the spectral analysis, we first checked the membership of the observed stars by using the \textit{Gaia} DR3 PM \citep{gaia_16,gaia_23}, which show a distribution peaked at $\mu_{\alpha}\text{cos}\delta = -5.403$, $\mu_{\delta} = -7.431$ mas yr$^{-1}$ \citep[see][]{vasiliev_21} and with a dispersion $\sigma_{\rm PM}\sim 0.5$ mas yr$^{-1}$ in the direction of Liller 1 (see also AG24). Hence, by assuming that all the stars with PMs within $3\times \sigma_{\rm PM}$ from the systemic value belong to the system, we conclude that 21 spectroscopic targets are likely members of Liller 1, and only star \#57 is a field interloper (see Figure~\ref{fig1}). The latter has therefore been excluded from the subsequent spectral analysis. The position of the 21 members stars in the near-infrared colour-magnitude diagram (CMD), and on the plane of the sky with respect to the centre of the system is shown in  Figure \ref{fig2}. All the observed stars are luminous giants located in the innermost 2 arcmin.

The surface temperature (T$_{\rm eff}$) and gravity ($\log g$) of the spectroscopic targets have been estimated photometrically from the projection of each star onto the closest isochrone in the CMD. Two isochrones \citep{bressan_12} properly matching the old and metal-poor, and the young and metal-rich components of Liller 1 have been considered:  one with an age of 12 Gyr and a metallicity [Fe/H]$=-0.3$, the other with and age of 2 Gyr and  [Fe/H]$=+0.3$, respectively. The comparison has been performed in the differential reddening corrected CMD, adopting a distance modulus (m-M)$_0$=14.65 and an average colour excess E(B-V)=4.52 \citep[see][]{Pallanca_21,ferraro_21}. 
In particular, we first projected all the stars on the isochrone with 12 Gyr and a metallicity of -0.3, representative of the bulk of the Liller~1 population, and deriving corresponding T$_{\rm eff}$ and log(g) values. We then used synthetic spectra with these stellar parameters to derive a first estimate of the metallicity. For those stars that turned out to have a different metallicity, we repeated the procedure, by projecting them on the younger, metal-rich isochrone, deriving new estimates for T$_{\rm eff}$, log(g) and then, metallicity.
The resulting values of T$_{\rm eff}$ and $\log g$ have then been fine-tuned by requiring the simultaneous fit of the observed OH and CO molecular lines and bandheads.
Low temperatures (in the 3300-3700 K range, with uncertainties of $\pm 100$ K), and surface gravities $\log g =0.5 \pm 0.3$ dex have been estimated. The adopted uncertainties of $\pm$100~K and $\pm$0.3 dex for the derived temperatures and gravities, respectively, also account for some possible degeneracy in identifying of the best-fit isochrone. In addition, for all the observed stars, a microturbulence velocity of $2 \pm 0.3$ km s$^{-1}$, which is typical of bulge giant stars with similar temperatures and metallicities, has been safely adopted.

The synthetic spectra used to measure the RVs and chemical abundances of the target stars have been computed by adopting the list of atomic lines from the VALD3 compilation \citealt{Ryabchikova_15}), molecular lines from the website of B. Plez\footnote{\url{https://www.lupm.in2p3.fr/users/plez/}}, MARCS models atmospheres (\citealt{gustafsson_08}) and the radiative transfer code TURBOSPECTRUM \citep{alvarez_98,plez_12}. 
We generated multiple grids of synthetic spectra, with fixed stellar parameters (appropriate to each star), and varying the metallicity from $-1.0$ dex to $+0.5$ dex, in steps of 0.1 dex, with both solar-scaled and some enhancement of [$\alpha$/Fe] and [N/Fe] and corresponding depletion of [C/Fe] for a proper computation of the molecular equilibria.
Solar-scaled [X/Fe] values have been adopted for the other elements.
For an optimal matching of the observed line profile broadening, the synthetic spectra have been convoluted with a Gaussian function at R$\approx$25,000 spectral resolution and resampled in pixels of $\approx$0.2~\AA, as the NIRSpec spectra. Such an instrumental broadening amounts to 12 km~s$^{-1}$. This is dominant with respect to any possible intrinsic broadening due, e.g., to macroturbulence and rotation, which typically do not exceeds 10 km~s$^{-1}$ in giant stars.

\section{Results} 
\label{results}
\begin{table*}[!t]
\caption{IDs, T$_{\rm eff}$, RVs and chemical abundances for the observed stars in Liller~1.}
\label{tab2}
%\centering
%\resizebox{15cm}{!}{
\scriptsize
\setlength{\tabcolsep}{5.25pt}
\renewcommand{\arraystretch}{1.4}
%\rotatebox{90}{
\begin{tabular}{|c|c|c|c|c|c|c|c|c|c|c|c|c|c|}
\hline\hline
ID & T$_{\text{eff}}$ & RV & [Fe/H] & [C/H] & [N/H] & [O/H] & [Mg/H] & [Al/H] & [Si/H] & [Ca/H] & [Ti/H] & [V/H] & $^{12}$C/$^{13}$C \\
\hline
\hline
 & [K] & km s$^{-1}$& 7.50$^{**}$ & 8.56$^{**}$ & 8.77$^{**}$ & 6.29$^{**}$ & 7.55$^{**}$ & 6.43$^{**}$ & 7.59$^{**}$ & 6.37$^{**}$ & 4.94$^{**}$ & 3.89$^{**}$ & 89$^{**}$ \\
\hline
1   &   3400  &   58  &  -0.03$\pm$0.04  &  -0.36$\pm$0.05  &   0.48$\pm$0.10  &   0.09$\pm$0.01  &   0.10$\pm$0.01  &   0.00$\pm$0.10  &   0.02$\pm$0.10  &  -0.01$\pm$0.03  &   0.02$\pm$0.10  &  -0.19$\pm$0.10  &  11.7$\pm$1  \\
3   &   3300  &   79  &  -0.08$\pm$0.05  &  -0.24$\pm$0.08  &   0.48$\pm$0.10  &   0.13$\pm$0.10  &   0.08$\pm$0.10  &   0.04$\pm$0.10  &   0.11$\pm$0.10  &   0.05$\pm$0.07  &   0.20$\pm$0.10  &  -0.04$\pm$0.10  &  12.5$\pm$1  \\
5   &   3400  &   62  &   0.25$\pm$0.05  &  -0.12$\pm$0.06  &   0.71$\pm$0.10  &   0.19$\pm$0.09  &   0.27$\pm$0.12  &   0.22$\pm$0.10  &   0.24$\pm$0.10  &   0.34$\pm$0.11  &   0.31$\pm$0.10  &   0.09$\pm$0.10  &   9.8$\pm$1  \\
6   &   3400  &  107  &  -0.14$\pm$0.08  &  -0.18$\pm$0.08  &   0.36$\pm$0.10  &   0.17$\pm$0.01  &   0.01$\pm$0.10  &  -0.04$\pm$0.10  &   0.15$\pm$0.10  &  -0.02$\pm$0.11  &   0.16$\pm$0.10  &  -0.16$\pm$0.10  &   8.7$\pm$1  \\
12  &   3400  &   60  &  -0.21$\pm$0.04  &  -0.40$\pm$0.06  &   0.55$\pm$0.10  &   0.18$\pm$0.04  &   0.22$\pm$0.10  &   0.02$\pm$0.10  &   0.17$\pm$0.10  &   0.22$\pm$0.09  &   0.20$\pm$0.10  &  -0.31$\pm$0.10  &   7.8$\pm$1  \\
14  &   3400  &   71  &   0.24$\pm$0.03  &  -0.15$\pm$0.07  &   0.64$\pm$0.12  &   0.26$\pm$0.04  &   0.28$\pm$0.10  &   0.27$\pm$0.10  &   0.25$\pm$0.10  &   0.30$\pm$0.04  &   0.26$\pm$0.10  &   0.17$\pm$0.10  &  12.8$\pm$1  \\
20* &   3400  &   52  &  -0.27$\pm$0.04  &  -0.34$\pm$0.07  &   0.46$\pm$0.10  &   0.30$\pm$0.10  &   0.10$\pm$0.10  &   0.15$\pm$0.08  &  -0.07$\pm$0.10  &   0.20$\pm$0.10  &   0.28$\pm$0.10  &  -0.08$\pm$0.10  &   6.8$\pm$1  \\
23  &   3400  &   68  &   0.36$\pm$0.04  &  -0.09$\pm$0.06  &   0.86$\pm$0.06  &   0.40$\pm$0.10  &   0.45$\pm$0.10  &   0.43$\pm$0.10  &   0.33$\pm$0.10  &   0.38$\pm$0.06  &   0.46$\pm$0.10  &   0.26$\pm$0.10  &   9.2$\pm$1  \\
24* &   3400  &   86  &   0.27$\pm$0.04  &  -0.08$\pm$0.05  &   0.78$\pm$0.10  &   0.32$\pm$0.02  &   0.31$\pm$0.10  &   0.34$\pm$0.05  &   0.26$\pm$0.10  &   0.32$\pm$0.02  &   0.38$\pm$0.10  &   0.30$\pm$0.10  &   8.0$\pm$1  \\
27* &   3400  &   69  &  -0.40$\pm$0.04  &  -0.72$\pm$0.06  &   0.29$\pm$0.00  &  -0.02$\pm$0.03  &   0.09$\pm$0.10  &  -0.07$\pm$0.10  &  -0.11$\pm$0.10  &  -0.01$\pm$0.07  &   0.11$\pm$0.10  &  -0.44$\pm$0.10  &  10.0$\pm$1  \\
31* &   3400  &   60  &  -0.28$\pm$0.03  &  -0.88$\pm$0.08  &   0.33$\pm$0.06  &   0.06$\pm$0.06  &   0.01$\pm$0.10  &   0.01$\pm$0.10  &  -0.06$\pm$0.10  &   0.02$\pm$0.07  &   0.26$\pm$0.10  &  -0.29$\pm$0.10  &   9.6$\pm$1  \\
35* &   3400  &   86  &   0.23$\pm$0.04  &  -0.14$\pm$0.03  &   0.80$\pm$0.09  &   0.28$\pm$0.01  &   0.24$\pm$0.10  &   0.34$\pm$0.10  &   0.27$\pm$0.10  &   0.24$\pm$0.01  &   0.40$\pm$0.10  &   0.25$\pm$0.10  &  11.5$\pm$1  \\
37* &   3400  &   61  &  -0.34$\pm$0.08  &  -0.51$\pm$0.06  &   0.20$\pm$0.08  &   0.16$\pm$0.08  &   0.06$\pm$0.10  &  -0.02$\pm$0.10  &  -0.06$\pm$0.10  &   0.11$\pm$0.10  &   0.13$\pm$0.10  &  -0.37$\pm$0.10  &   8.0$\pm$1  \\
39* &   3400  &   55  &  -0.33$\pm$0.03  &  -0.80$\pm$0.03  &   0.39$\pm$0.03  &   0.10$\pm$0.01  &   0.16$\pm$0.10  &  -0.09$\pm$0.10  &  -0.05$\pm$0.10  &  -0.06$\pm$0.08  &   0.18$\pm$0.10  &  -0.32$\pm$0.10  &   9.9$\pm$1  \\
41  &   3400  &   85  &  -0.37$\pm$0.04  &  -0.44$\pm$0.06  &   0.16$\pm$0.09  &   0.08$\pm$0.11  &   0.06$\pm$0.02  &  -0.06$\pm$0.10  &   0.12$\pm$0.10  &  -0.06$\pm$0.07  &   0.06$\pm$0.10  &  -0.46$\pm$0.10  &   8.6$\pm$1  \\
63  &   3400  &   66  &  -0.32$\pm$0.02  &  -0.66$\pm$0.05  &   0.45$\pm$0.10  &   0.08$\pm$0.10  &   0.07$\pm$0.10  &  -0.04$\pm$0.10  &  -0.04$\pm$0.10  &   0.02$\pm$0.06  &  -0.05$\pm$0.10  &  -0.40$\pm$0.10  &  12.7$\pm$1  \\
67  &   3400  &   76  &  -0.34$\pm$0.03  &  -0.48$\pm$0.06  &   0.29$\pm$0.10  &   0.13$\pm$0.10  &  -0.07$\pm$0.10  &  -0.07$\pm$0.10  &   0.09$\pm$0.10  &  -0.10$\pm$0.05  &   0.13$\pm$0.10  &  -0.43$\pm$0.10  &  12.8$\pm$1  \\
70  &   3400  &   45  &  -0.48$\pm$0.04  &  -0.62$\pm$0.05  &   0.21$\pm$0.10  &  -0.11$\pm$0.10  &   0.13$\pm$0.10  &  -0.01$\pm$0.10  &  -0.04$\pm$0.10  &  -0.08$\pm$0.10  &   0.09$\pm$0.10  &  -0.42$\pm$0.10  &  10.0$\pm$1  \\
74* &   3500  &   70  &   0.29$\pm$0.06  &  -0.22$\pm$0.06  &   0.97$\pm$0.10  &   0.31$\pm$0.10  &   0.37$\pm$0.10  &   0.37$\pm$0.10  &   0.26$\pm$0.10  &   0.32$\pm$0.06  &   0.36$\pm$0.10  &   0.15$\pm$0.10  &  13.3$\pm$1  \\
91  &   3600  &   62  &  -0.41$\pm$0.02  &  -1.12$\pm$0.08  &   0.07$\pm$0.03  &   0.00$\pm$0.09  &  -0.04$\pm$0.01  &  -0.11$\pm$0.10  &  -0.14$\pm$0.10  &   0.00$\pm$0.01  &   0.01$\pm$0.10  &  -0.38$\pm$0.10  &  14.1$\pm$1  \\
120*&   3700  &   68  &  -0.33$\pm$0.04  &  -0.86$\pm$0.06  &   0.47$\pm$0.10  &  -0.06$\pm$0.03  &   0.12$\pm$0.10  &   0.14$\pm$0.10  &  -0.03$\pm$0.10  &  -0.01$\pm$0.08  &   0.03$\pm$0.10  &  -0.35$\pm$0.10  &  11.1$\pm$1  \\
\hline\hline
\end{tabular}

\vspace{0.15cm}
Notes: \\
$^*$Stars in common with AG24.\\
$^{**}$Adopted solar abundances for the measured chemical elements are from \citet{magg_22} and they are reported in the header of each element abundance column.
\end{table*}
\subsection{Radial velocities}
The RVs of the 21 member stars have been measured by means of cross-correlation between the observed and the synthetic spectra. The resulting values (see Table \ref{tab2}) range between 45 and 110 km~s$^{-1}$ range, with uncertainties smaller than 1 km s$^{-1}$, and fully consistent (within 1 km s$^{-1}$) with the measurements quoted by AG24 for the nine stars in common. The sample has mean value of 68.8$\pm$3.1 km~s$^{-1}$, which is in very good agreement with the average RVs quoted in \citet[][67.9$\pm$0.8 km~s$^{-1}$]{crociati_23} and in AG24 (66.6$\pm$2.7 km~s$^{-1}$). The measured RV dispersion is 14.1$\pm$2.2 km~s$^{-1}$, well consistent with the line-of-sight velocity dispersion of $\approx$13~km~s$^{-1}$ measured at about $100\arcsec$ from the center of Liller~1 by \citet{baumgardt_18}\footnote{Fundamental parameters of Galactic globular clusters, \url{https://people.smp.uq.edu.au/HolgerBaumgardt/globular/}}. All the 21 stars have RVs within 3$\sigma$ from the systemic one, thus confirming their membership from PMs.
\begin{figure}
    \centering
    \includegraphics[width=0.97\columnwidth]{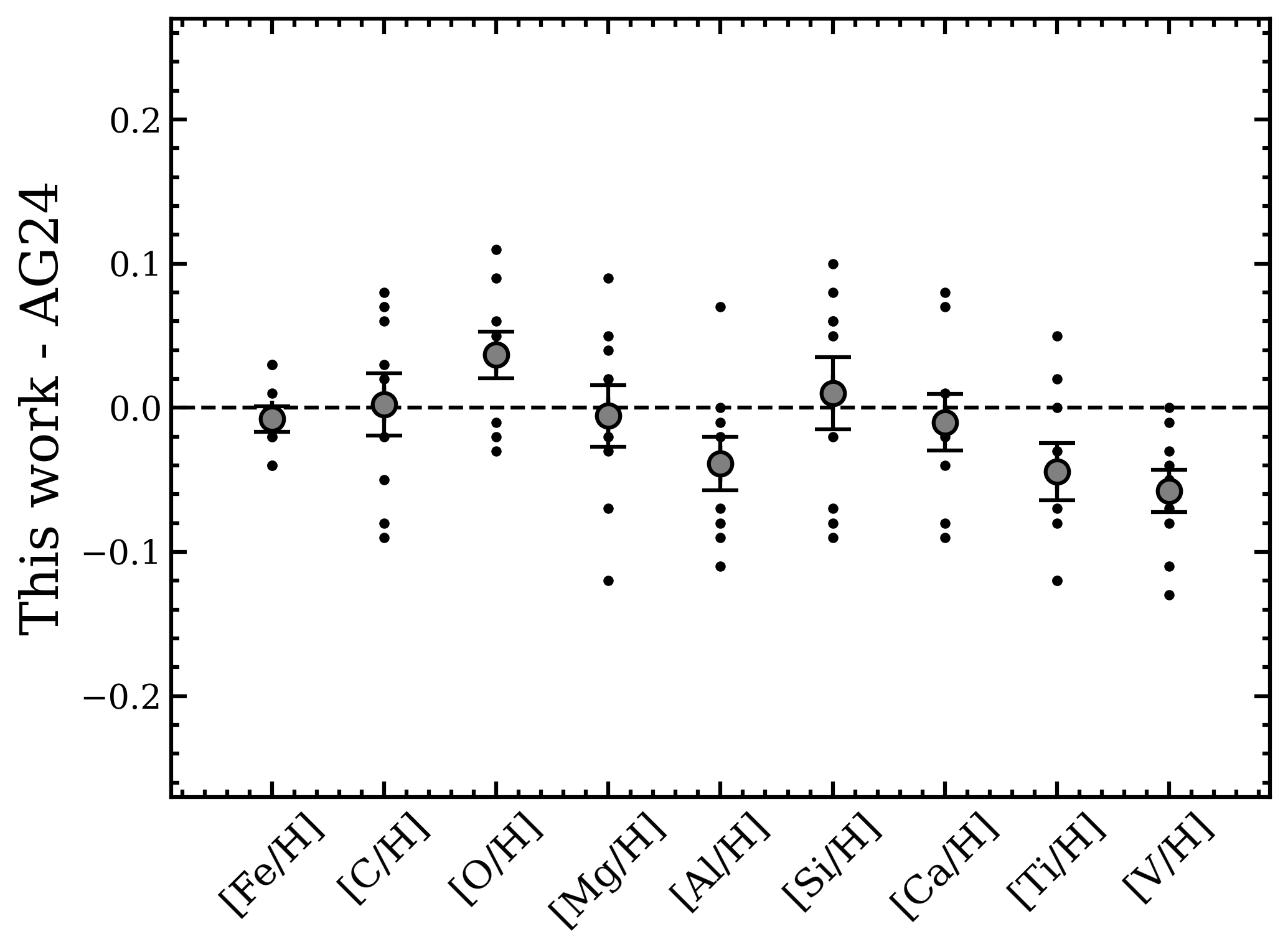}
    \caption{Abundance differences between this work and AG24 for the stars and the chemical elements in common. Gray dots refer to the measurements obtained for each individual star, while the black filled circles are the average differences along with their associated errors. The dashed  horizontal line marks the zero difference.}
    \label{dif}
\end{figure}
\begin{figure}
    \centering
    \includegraphics[width=\columnwidth]{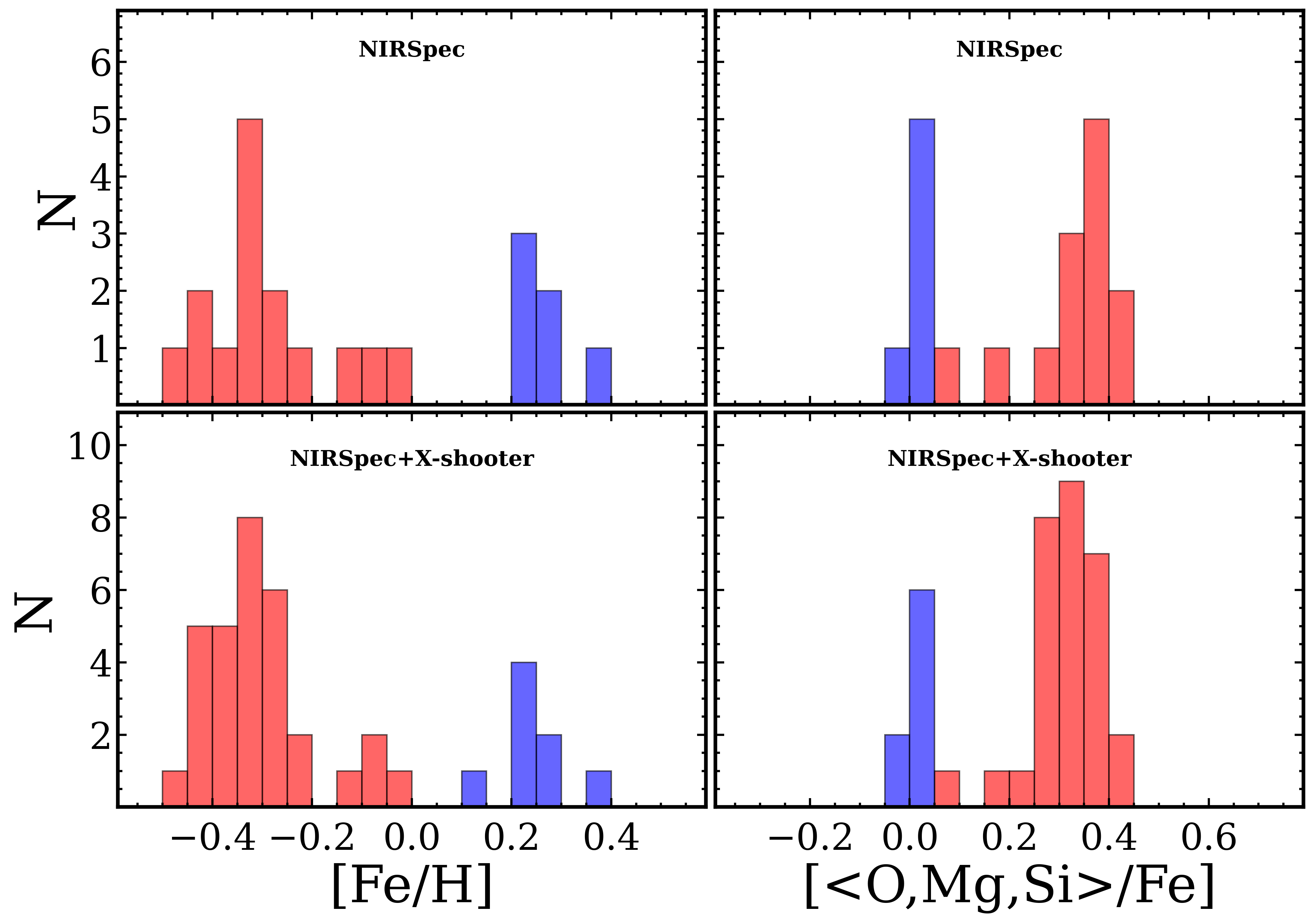}
    \caption{Distribution of the [Fe/H] and [$<$O,Mg,Si$>$/Fe] abundance measurements (left and right panels, respectively) 
    obtained in the present work from the analysis of NIRSpec spectra alone (top panel), and in combination with the sample AG24 sample (bottom panels).}
    \label{hist}
\end{figure}
\begin{figure*}
    \centering
    \includegraphics[width=0.95\textwidth]{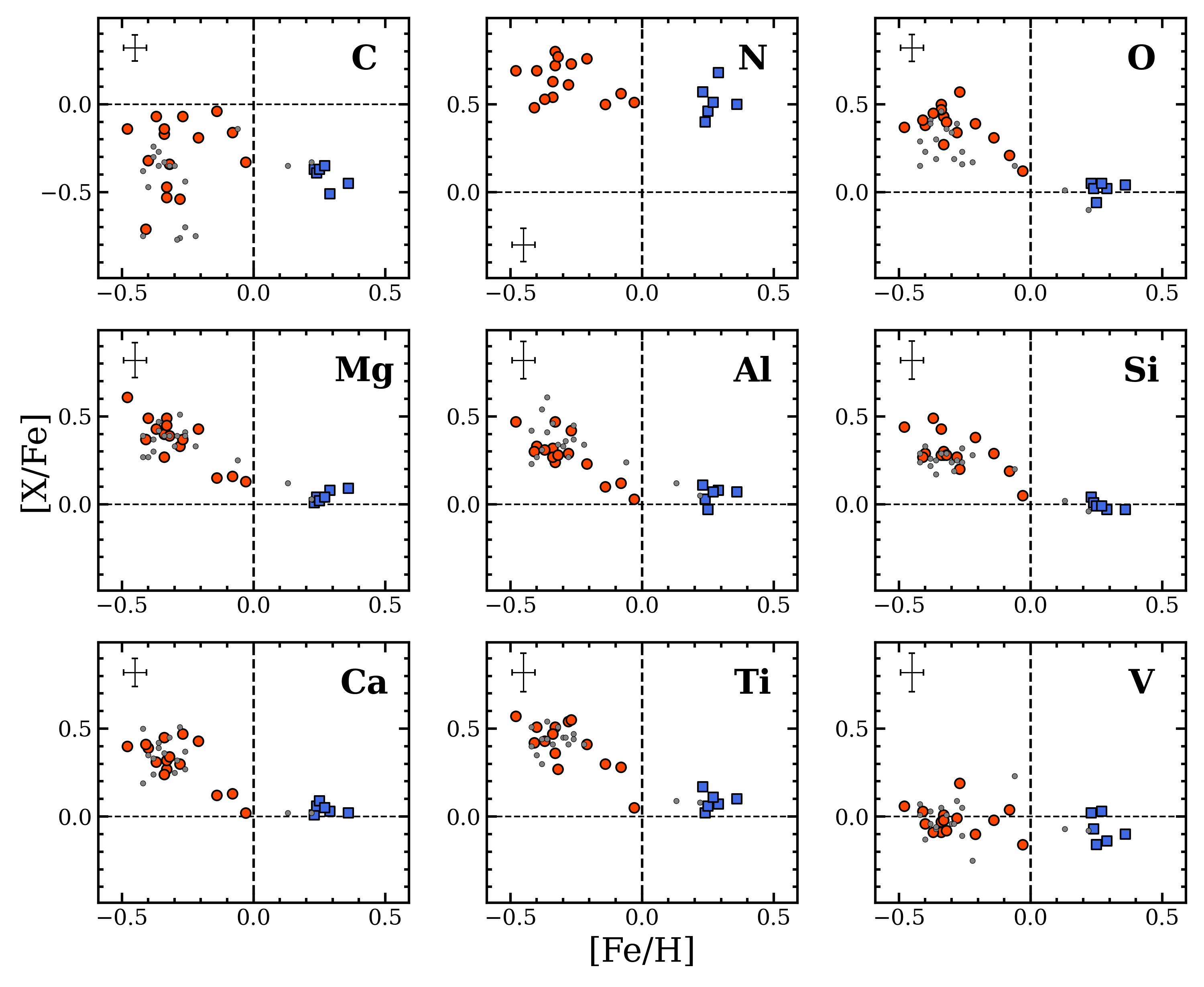}
    \caption{Behavior of [C/Fe], [N/Fe], [O/Fe], [Mg/Fe], [Al/Fe], [Si/Fe], [Ca/Fe], [Ti/Fe], and [V/Fe] as a function of [Fe/H] for the metal-poor (red filled circles) and metal-rich (blue filled squares) subpopulations of Liller~1. The gray dots are the stars of AG24 not in common with the NIRSpec sample. The typical error bars are reported in the left corner of each panel.
    The dashed vertical and horizontal lines denote the corresponding zero [Fe/H] and [X/Fe] values.}
    \label{ratio}
\end{figure*}
\begin{figure}
    \centering
    \includegraphics[width=\columnwidth]{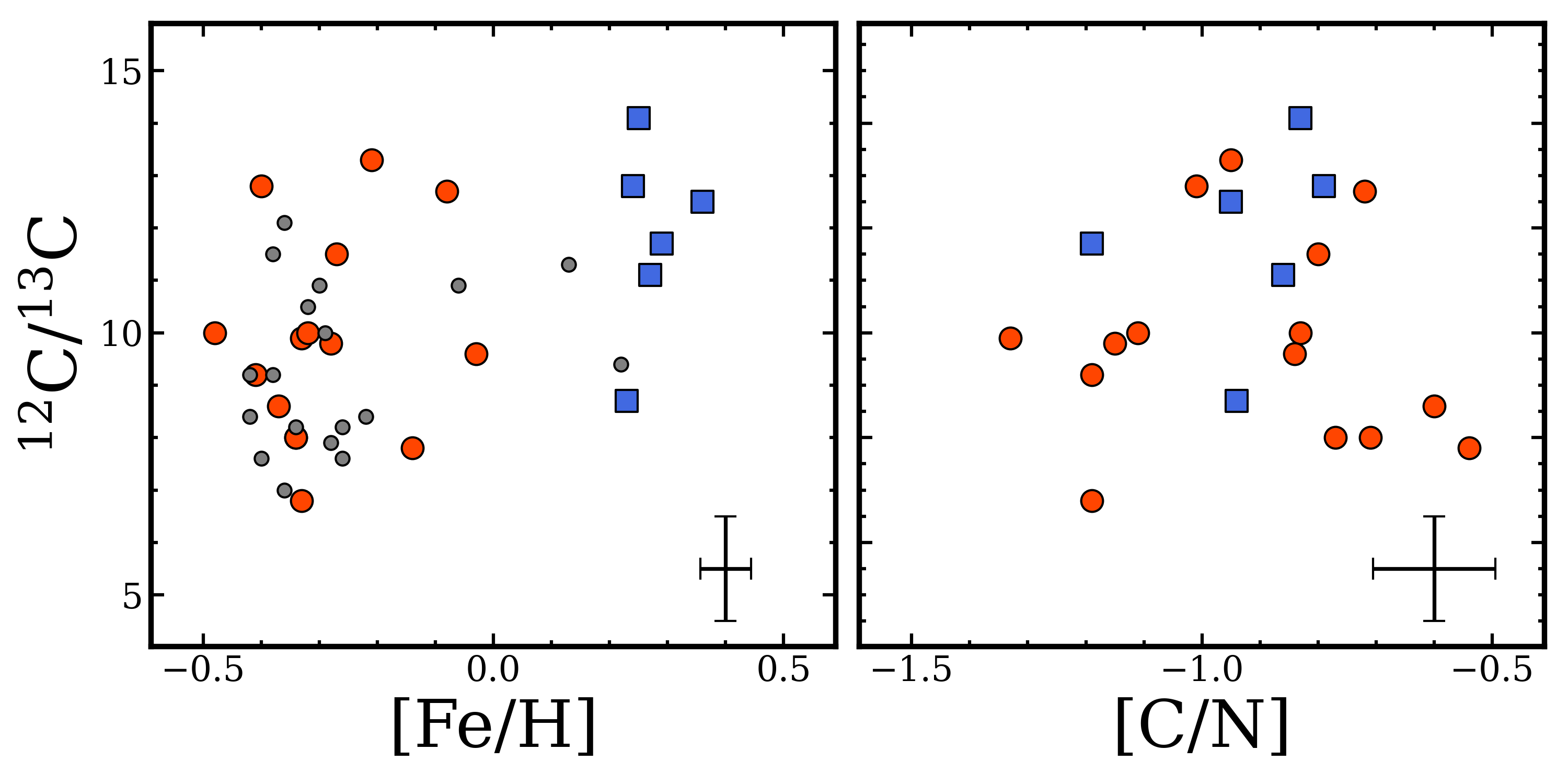}
    \caption{$^{12}$C/$^{13}$C isotopic ratio as a function of [Fe/H] (left panels), and 
    [C/N] (right panel). Typical error bars are reported in the bottom-right corner of each panel. Symbols are as in Fig.~\ref{ratio}.}
    \label{CNC13}
\end{figure}
\subsection{Chemical abundances and abundance ratios}
\label{abu}
 The chemical abundances of Fe, CNO, Ca, Si, Mg, Ti, Al and V have been determined via spectral synthesis around each line of interest. For each measured star and chemical element, Table~\ref{tab2} lists the average abundance (if two or more lines are used) and corresponding error, i.e. the standard deviation divided by the square root of the number of lines used, or a conservative value of  0.1 dex if only one line is used. 
 While chemical abundances of Fe, Ca, Si, Mg, Ti, Al and V have been determined from atomic lines, those of N and O have been derived from CN and OH molecular transitions and those of $^{12}$C and $^{13}$C from the 2$^{\rm nd}$ overtone CO molecular bandheads. Indeed, as discussed in \citet[][and references therein]{fanelli21}, reliable C abundances from CO can be determined either from individual roto-vibrational transitions and/or from the bandheads.

At the low effective temperatures of these targets, the estimated uncertainties on the stellar parameters (see previous section) have an overall impact of approximately 0.10-0.15 dex  on the derived abundances. However, it is worth mentioning that this global uncertainty can be regarded as mostly systematic, thus being essentially canceled out when computing abundance ratios, as well as when abundance differences among the Liller~1 stars are considered.

Nine targets in our NIRSpec sample are in common with the study of AG24. The abundance differences for the chemical elements in common are symmetrically distributed around the zero value, typically well within 0.1~dex and comparable to the measurement errors (see Figure \ref{dif}). The average differences are always within 0.05 dex, indicating that the two sets of abundance measurements can be merged without introducing major systematic biases.

The distribution of the inferred [Fe/H] values for the  21 likely member stars of Liller~1 is shown in Figure \ref{hist} (left panel). It shows a major, sub-solar component counting 15 stars at an average [Fe/H]=$-0.29\pm 0.03$ and 1$\sigma$ dispersion of 0.13$\pm$0.02 (which exceeds the measurement errors), and a super-solar one at an average [Fe/H]=$+0.27 \pm 0.02$ and $1 \sigma =0.05 \pm 0.01$, comprising 6 stars. 
Interestingly, such a bi-modal distribution is fully consistent with the one found by \citet{crociati_23} and AG24, and it agrees with the prediction obtained from the reconstructed star formation history of the system \citep{dalessandro_22}.
The stars belonging to the two components are highlighted with different shape and colours (red filled circles and blue filled squares for the sub- and super-solar components, respectively) in all the figures.

\begin{figure}
    \centering
    \includegraphics[width=0.985\linewidth]{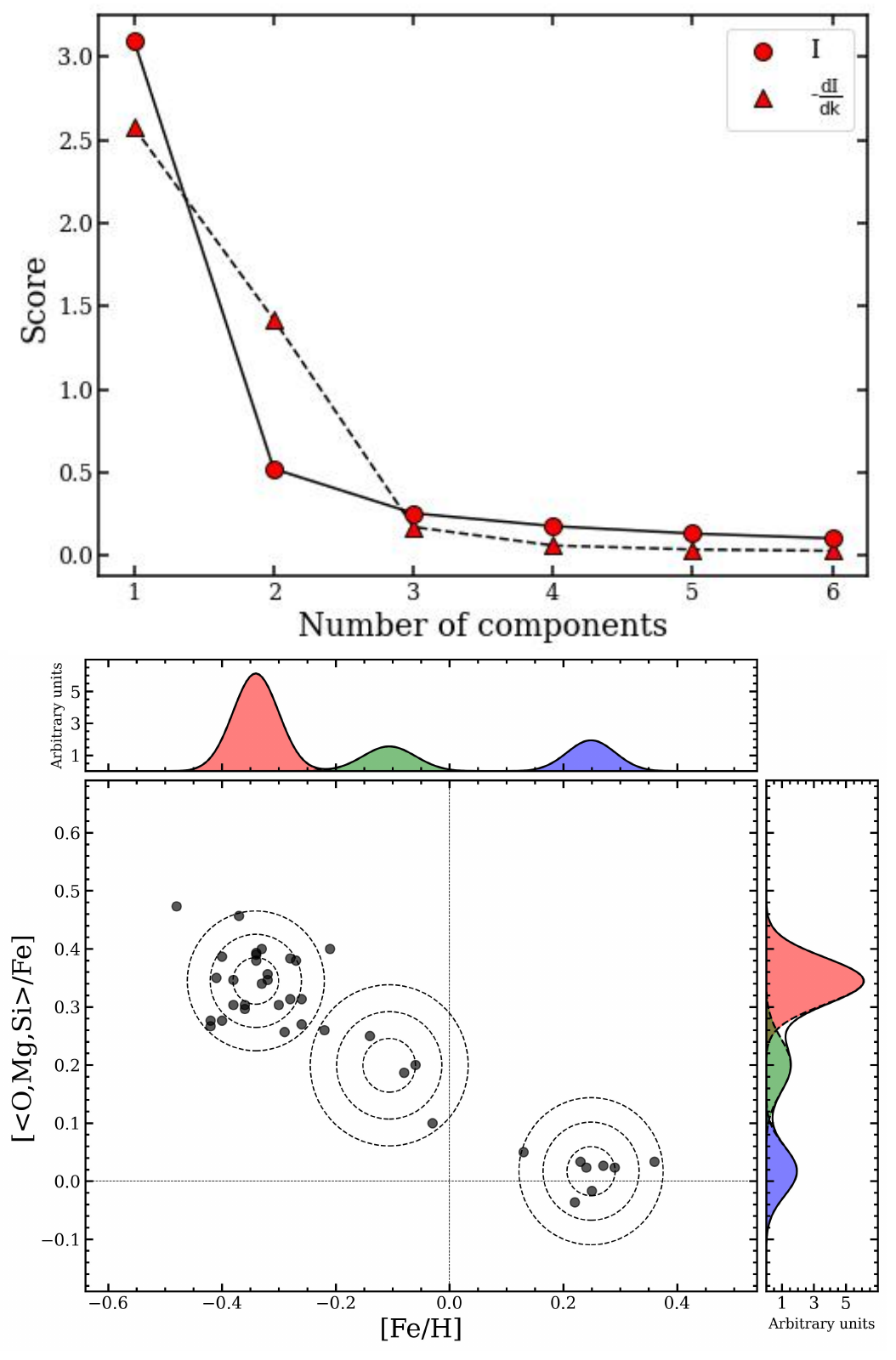}
    \caption{K-means clustering solution for the [$<$O,Mg,Si$>$/Fe] {\it vs} [Fe/H] distribution of Liller~1. 
     Bottom panel: distribution for the observed 39 stars (black filled circles) and best fit solution with three Gaussian components and 1,2,3 $\sigma$ contours, where $\sigma$ is the standard deviation. Top panel: inertia parameter I(k) (red filled circles and solid line) and its derivative (red triangles and dashed line) as a function of $K$, the total number of components.}
    \label{kmeans}
\end{figure}

\begin{figure}
    \centering
    \includegraphics[width=1.011\linewidth]{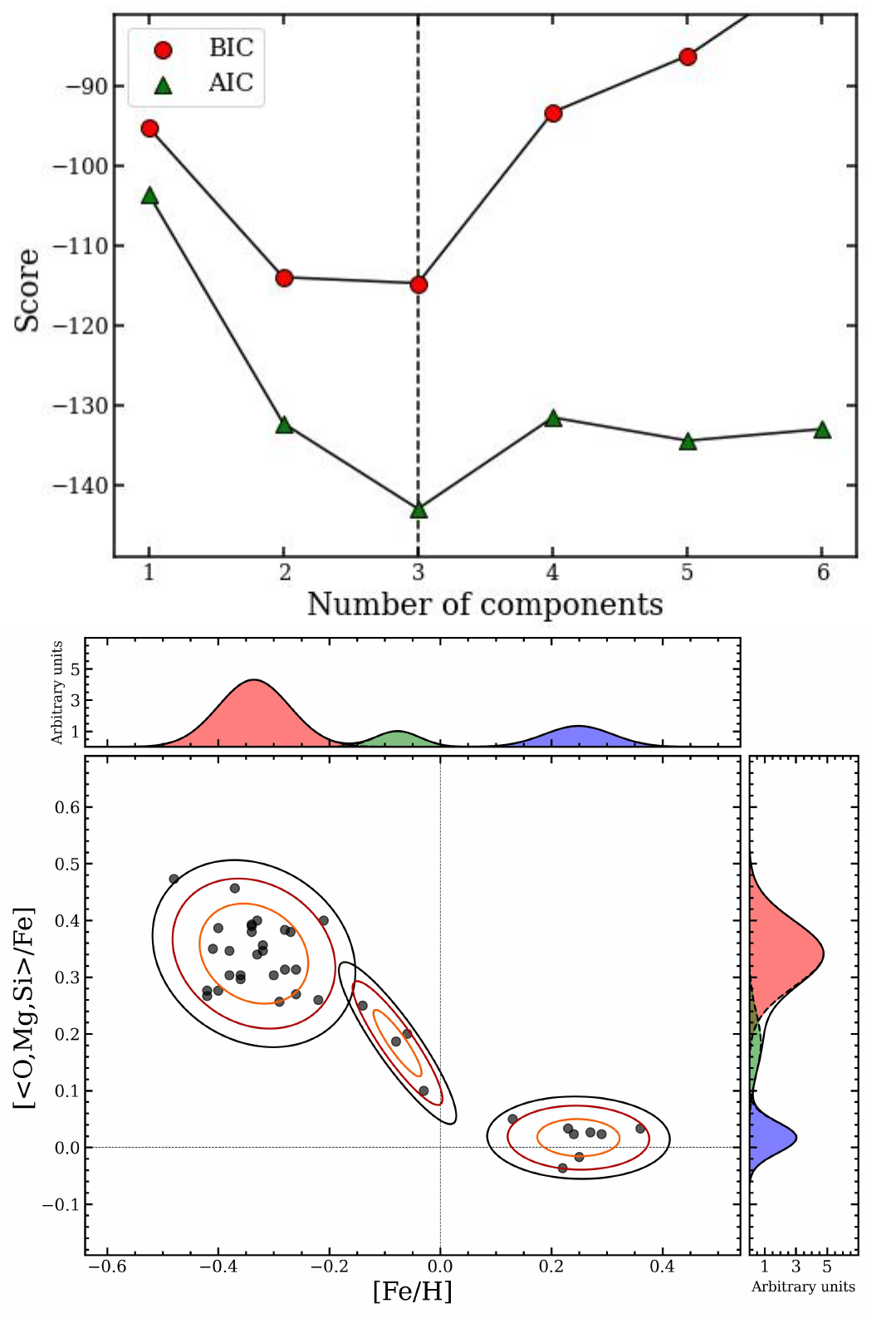}
    \caption{GMM solution for the [$<$O,Mg,Si$>$/Fe] {\it vs} [Fe/H] distribution of Liller~1.
    Bottom panel: distribution for the observed 39 stars (black filled circles) and best fit solution with three Gaussian components and 1,2,3 $\sigma$ contours, where $\sigma$ is the standard deviation of each component. Top panel: BIC (red filled circles) and AIC (green, big triangles) parameters as a function of the total number of Gaussian components.}
    \label{GMM}
\end{figure}

Concerning the other chemical elements measured in the NIRSpec spectra, Fig.~\ref{ratio} displays their [X/Fe] abundance ratios as a function of [Fe/H]. 
The metal-poor subpopulation shows some enhanced (on average by a factor of 2-3) [O/Fe], [Mg/Fe], [Al/Fe], [Si/Fe], [Ca/Fe] and [Ti/Fe], while the super-solar component has about solar-scaled values.
Interestingly, the few stars with intermediate iron content (in the -0.2$<$[Fe/H]$<$+0.2 dex range) show also somewhat intermediate enhancement in [O/Fe], [Mg/Fe], [Si/Fe], [Ca/Fe] and [Ti/Fe], in between the main metal-poor and metal-rich subcomponents.
The rather clear separation between the $\alpha$-enhanced, metal-poor population and the solar-scaled, metal-rich one can be appreciated in the right panel of Fig.~\ref{hist},which shows the average [<O,Si,Mg>/Fe] distribution for the observed stars.
Irrespective of their iron content, all the stars show enhancement of [N/Fe] up to a factor of about six, and about solar-scaled [V/Fe] values with small scatters.
[C/Fe] is depleted with respect to the solar-scaled value in all the stars, but at sub-solar [Fe/H] the scatter is significantly larger than at [Fe/H]$>0$, with an overall 1$\sigma$ dispersion exceeding 0.2 dex. We did not find any hint of Mg-Al or C-N anti-correlations, confirming the results of AG24 on the lack of chemical anomalies among light elements, also similarly to what found in Terzan~5 \citep{Origlia_11}.

Figure \ref{CNC13} shows the distributions of $^{12}$C/$^{13}$C as a function of [Fe/H] (left panel) and [C/N] (right panel).
Low $^{12}$C/$^{13}$C isotopic ratios (in the 5-15 range) with an average value of $\sim 10$ have been measured regardless the star iron content.
[C/N] values below -0.5 dex and as low as -1.4 dex have been obtained, with significantly larger scatter among the metal-poor than the metal-rich stars, thus reflecting the [C/Fe] distribution.
The inferred low [C/Fe] and $^{12}$C/$^{13}$C abundance ratios are the typical signature that mixing and extra-mixing processes occurred in the stellar interiors during the Post-Main Sequence evolution \citep[see, e.g.,][]{cha95,dw96,csb98,bs99}.

\begin{figure*}[!h]
    \centering
    \includegraphics[width=1\linewidth]{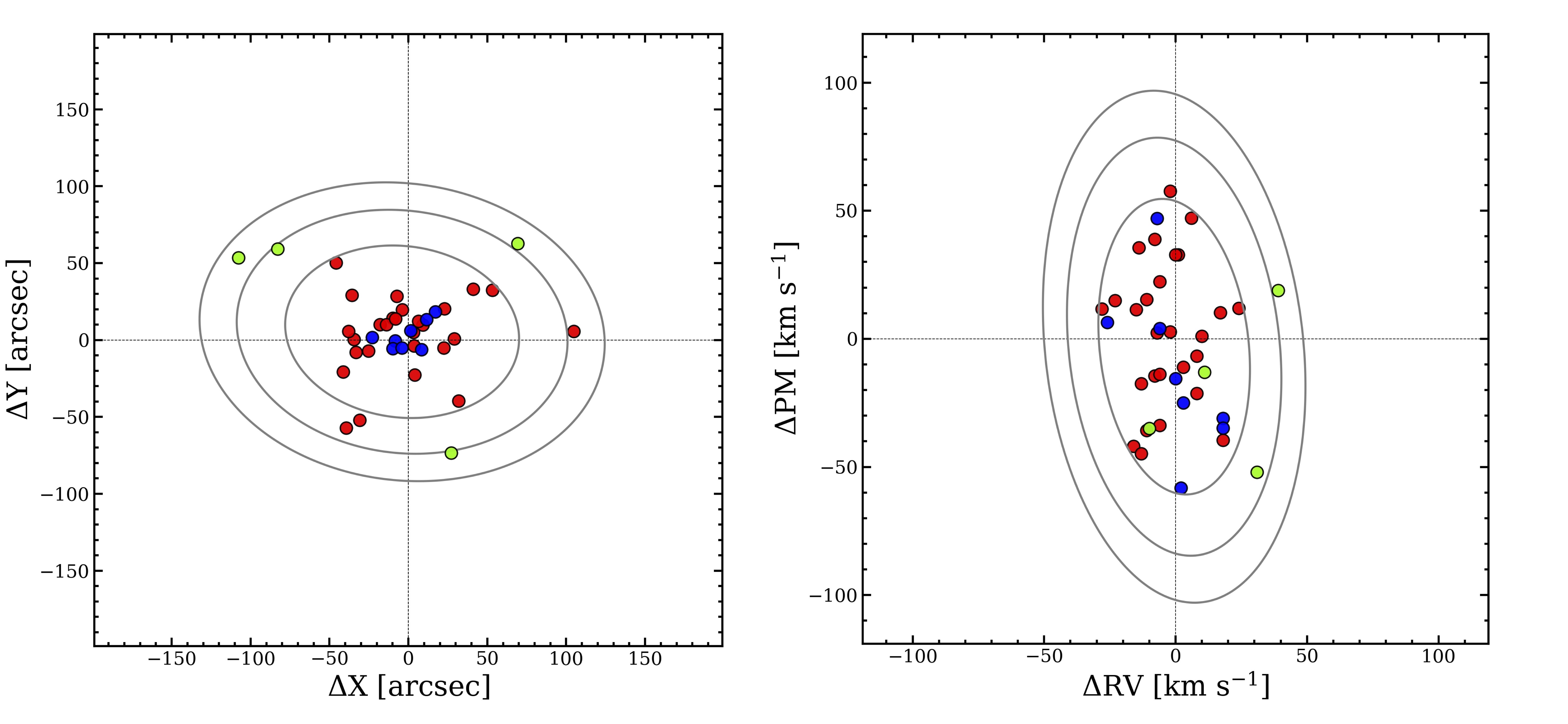}
    \caption{Spatial (in the RA and Dec plane referred to the system center, left panel) and kinematic (in the RV-PM plane referred to the systemic velocities, right panel) distributions of the 39 stars likely members of Liller~1, color-coded according to their metal content: red (metal-poor), green (metal-intermediate) and blue (metal-rich). The 1,2,3 $\sigma$ contour levels from the GMM analysis are also plotted.}
    \label{poskin}
\end{figure*}

The abundance ratio distributions obtained in this work fully match those of AG24 for the chemical elements in common.
Hence, we can merge the two data sets, obtaining a global sample counting 39 stars, for which abundances of Fe, C, O, Mg, Al, Si, Ca, Ti, V and the $^{12}$C/$^{13}$C isotopic ratio are available.

\section{A third, metal-intermediate component}
\label{disc}

The histograms shown in Fig.~\ref{hist} and the distributions of [$<$O,Mg,Si$>$/Fe] plotted in Fig.~\ref{ratio} suggest the possible presence of multiple (>2) components, although the low number statistics remains a somewhat limiting factor.
In order to  more quantitatively explore this hypothesis, we analyzed the  [$<$O,Mg,Si$>$/Fe] {\it vs} [Fe/H]
distribution of the 39 stars in the merged (this work and AG24) sample by means of two different statistical methods (see e.g. the {\tt scikit-learn}\footnote{\url{https://scikit-learn.org/stable/}} library), namely the \textit{K}-means clustering and the Gaussian Mixture Model (GMM). The Expectation-Maximization algorithm has been used to improve the fit to the data by iteratively varying the model parameters.

The K-means is a non-supervisioned clustering algorithm that provides the optimal number of components from the inertia parameter I defined as 
\begin{equation*}
   I(k) =  \sum_{j=1}^{k} \sum_{i=1}^{n} \delta_{i,j} \cdot \| x_i - \mu_j \|^2
\end{equation*}
where $k$ is the total number of \textit{j-th} 
components, \textit{n} is the total number of \textit{i} datapoints, $\mu_j$ the corresponding centroid, $x_i$ is the 2D coordinate of each \textit{i-th} datapoint, $\delta_{i,j}$ is a $\delta$ function which assesses the membership of the \textit{i-th} point to the \textit{j-th} component, and $\| x_i - \mu_j \|$ is the euclidean distance.
The optimal number of components is defined by the "elbow" in the inertia {\it vs} $k$ diagram, at which the inertia and its derivative sharply level off.
Figure~\ref{kmeans} shows the inferred K-means optimal solution for the 
[$<$O,Mg,Si$>$/Fe] {\it vs} [Fe/H] distribution of Liller~1. 
We explored solutions with a total number of components between 1 and 6, and we found that the optimal one turns out to be $k$=3. 
\begin{table}[!t]
\setlength{\tabcolsep}{3pt}
\centering
\caption{Liller~1 components (metal-poor, metal-intermediate and metal-rich) as derived from a GMM decomposition of the observed [$<$O,Mg,Si$>$/Fe] \textit{vs} [Fe/H] distribution.}
\tiny
\begin{tabular}{@{}lrcccr@{}}
\toprule
Component & <[Fe/H]> & $\sigma$ & [$<$O,Mg,Si$>$/Fe] & $\rm \sigma$ & N$_{star}$  \\ 
\midrule
Metal-poor    & -0.34$\pm$0.01 &0.06$\pm$0.01  & 0.34$\pm$0.01 &0.06$\pm$0.01 & 27 (69\%) \\
Metal-int  & -0.08$\pm$0.02 & 0.04$\pm$0.01 & 0.18$\pm$0.03 &0.05$\pm$0.01 & 4 (10\%) \\
Metal-rich    & 0.25$\pm$0.02 &0.06$\pm$0.02 & 0.02$\pm$0.01 &0.03$\pm$0.01 & 8 (21\%) \\
\bottomrule
\end{tabular}
\label{test}
\end{table}
Although the distribution of the two variables   ([$<$O,Mg,Si$>$/Fe] and [Fe/H]) separately is well reproduced, the model is unable to properly describe the 2D distribution of the stars in the diagram  because they are not uniformly and spherically distributed (see the bottom panel of Fig. \ref{kmeans}).

This suggests that an alternative approach (as the GMM) that allows  a  decomposition of an observed distribution into a given number of Gaussian components with their own mean, variance and rotation parameter is more appropriate.
As figures of merit for the selection of the best-fit solution with the GMM
we employed the Bayesian Information Criterion (BIC) and the Akaike Information Criterion (AIC). The BIC and the AIC are statistical methods that balance model complexity against fitting to the data. 
BIC is defined as $k\times {\rm ln}(n)-2ln(\hat{L})$, while AIC is defined as $2k-2ln(\hat{L})$, 
where \textit{k} is the number of parameters, \textit{n} the sample size, and $\hat{L}$ is the maximum likelihood of the model.
BIC increases with the number of model parameters and logarithmically with the sample size, while it logarithmically decreases with the maximum likelihood of the model. AIC has no direct dependence on the sample size and it primarily depends on the maximum likelihood of the model. 
The best-fit model is the one that minimizes both BIC and AIC.

We explored solutions with varying the number of Gaussian components from 1 to 6. The best-fit to the data is obtained with three components, as indicated by both the BIC and AIC parameters in the upper panel of Figure \ref{GMM}. The bottom panel of the same figure  shows the GMM optimal solution for the considered distribution, and Table~\ref{test} summarizes the parameters of the best-fit model.
In order to evaluate the statistical significance of the three component model with respect to the two component model, we also computed the  p-value by means of likelihood-ratio test and a chi-squared distribution with one degree of freedom. We find a p-value of $\approx$0.01, indicating that the three-component model has high statistical significance and is therefore the most appropriate for describing the observed distribution.

We finally investigated on a statistical ground the distribution of the global sample of 39 stars in the spatial and velocity planes. 
We applied a GMM to determine the optimal number of Gaussian components to fit the RA-Dec and RV-PM 2D distributions. Both the BIC and AIC are increasing with the number of Gaussian components used to fit the data, thus indicating that the best-fitting model is a single Gaussian. Fig.~\ref{poskin} shows the spatial and kinematic distributions of the 39 stars likely members of Liller~1, color-coded according to their metal content, and the 1,2,3 $\sigma$ contour levels from the GMM analysis. 
This test indicates that the 39 stars have spatial and velocity distributions consistent with  being members of the same stellar system.

\section{Discussion and conclusions}
\label{conclu}
This paper presents a high resolution spectroscopic study in the H-band of a significant sample of giant stars (21 objects), members of the bulge stellar system Liller~1, according to their 3D kinematics, and following the pioneering work by \citet{origlia_02} and the studies of \citet{crociati_23}, AG24) at lower spectral resolution. 
In particular, the nine stars in common between this study and AG24 have allowed to properly cross-check the inferred abundances for the chemical elements in common and to conclude that they are fully consistent within the errors. By merging the two samples, chemical abundances of Fe, C, O, Mg, Al, Si, Ca, Ti and V are thus available for 39 stars, likely members of Liller~1. 

From a statistical analysis of the [$\alpha$/Fe] vs [Fe/H] distribution of such a significantly larger (by 30\%) sample of stars compared to the AG24 one, 
we could provide the first chemical detection of a minor, third subcomponent with [Fe/H] and [$\alpha$/Fe] abundance ratios somewhat intermediate between the values characterizing the two main subpopulations, namely the one with sub-solar iron and [$\alpha$/Fe] enhancement that likely formed at an early epoch from a gas enriched by type II SNe and and the one with super-solar iron and solar-scaled [$\alpha$/Fe], that likely formed at much later epochs from a gas also enriched by type Ia SNe on a longer timescale and it is  more centrally concentrated. The three subcomponents have similar kinematic properties, consistent with belonging to the same stellar system.

Although more statistics is needed to firmly establish whether this third component is truly a distinct subpopulation or a pronounced tail toward higher metallicities of the main, metal-poor one, this finding represents an important tile in the puzzle of the origin of Liller~1. In fact it provides the observational signature that this stellar system cannot be explained as the simple merging of two genuine globulars \citep{khoperskov18,mastrobuono19,pfeffer21} or by a genuine globular accreting a giant molecular cloud \citep{mckenzie18,bastian_22}, since its age, metallicity and [$\alpha$/Fe] distributions are by far more complex, and characterized by multiple episodes of star formation likely accompanied by some self-enrichment.

High resolution spectroscopy of an additional large sample of giant stars is currently ongoing at the ESO-VLT with the spectrograph CRIRES+ in the contest of the survey BulCO ("The Bulge Cluster Origin") a Large Programme specifically aimed at settling the origin of the most massive stellar clusters in the Galactic Bulge. This sample will allow to increase the statistics of the distribution, especially in the least populated region of the [$\alpha$/Fe]-[Fe/H] diagram around solar metallicity, thus finally constraining the details of the complex star formation and chemical evolution history of this fascinating stellar system.

\begin{acknowledgements}
CF and LO acknowledge the financial support by INAF within the VLT-MOONS project. CF would like to thank R. Pascale for useful discussions. This work is part of the project Cosmic-Lab at the Physics and Astronomy Department “A. Righi” of the Bologna University (\url{http:// www.cosmic-lab.eu/ Cosmic-Lab/Home.html}). 
\end{acknowledgements}
%
%
% - use BibTeX with the regular commands:
\bibliographystyle{aa} % style aa.bst
\bibliography{biblio} % your references Yourfile.bib
%
% - join the .bib files when you upload your source files
%-------------------------------------------------------------------
%
%panel con 
\end{document}